\documentclass[aps,amssymb,amsmath,amsfonts,twocolumn,pra,superscriptaddress]{revtex4}
\usepackage{amssymb}
\usepackage{diagbox} 
\usepackage{amsmath}
\usepackage[english]{babel}
\usepackage[latin1]{inputenc}
\usepackage[usenames, dvipsnames]{color}
\usepackage{graphicx}
\usepackage{gensymb}

\begin{document}
\title{Operational approach to Bell inequalities: application to qutrits}
\author{Daniel Alsina}
\affiliation{Dept. F\'isica Qu\`antica i Astrof\'isica, Universitat de Barcelona, Av. Diagonal 647, 08028 Barcelona, Spain.}
\author{Alba Cervera}
\affiliation{Dept. F\'isica Qu\`antica i Astrof\'isica, Universitat de Barcelona, Av. Diagonal 647, 08028 Barcelona, Spain.}
\author{Dardo Goyeneche}
\affiliation{Faculty of Physics, Warsaw University, Pasteura 5, 02-093 Warsaw, Poland}
\affiliation{Faculty of Applied Physics and Mathematics, Technical University of Gda\'{n}sk, 80-233 Gda\'{n}sk, Poland}
\affiliation{Institute of Physics, Jagiellonian University, ul. Reymonta 4, 30-059 Krak\'{o}w, Poland}
\author{Jos\'e~I.~Latorre}
\affiliation{Dept. F\'isica Qu\`antica i Astrof\'isica, Universitat de Barcelona, Av. Diagonal 647, 08028 Barcelona, Spain.}
\author{Karol {\.Z}yczkowski}
\affiliation{Institute of Physics, Jagiellonian University, ul. Reymonta 4, 30-059 Krak\'{o}w, Poland}
\affiliation{Center for Theoretical Physics, Polish Academy of Sciences, al. Lotnik\'{o}w 32/46, 02-668 Warszawa, Poland\vspace{0.5cm}}

\date{August 11, 2016}

\begin{abstract}
In this work we develop two methods to construct Bell inequalities for multipartite systems. By considering non-hermitian operators we study Bell inequalities for the cases of three settings, three outcomes and three to six parties. The maximal value achieved in the framework of quantum theory is computed for subsystems with three levels each. The other technique, based on a mapping from pure entangled  states to Bell operators, allows us to construct further multipartite Bell inequalities. As a consequence, we reproduce some known results in a novel way and find some multipartite Bell inequalities for systems having three settings and three outcomes per party.
\end{abstract}

\maketitle

\section{Introduction}
In 1964, Bell introduced an inequality that provided a tool to discern between quantum non-locality and any local theory of hidden variables \cite{Bell}. A new Bell inequality was proposed in the 1969 CHSH paper \cite{CHSH}, which was simpler and easier to test experimentally.
It placed constraints on expected values of measurements of correlations of two outcomes with two settings per observer. Experimenters quickly began to test the inequality, and by 1982 there was already 
a strong evidence that local hidden variable theories were being ruled out \cite{Aspect}. However, the question of loopholes remained alive: hypotheses on the experimental setting that were taken for granted while computing the expectation values and that were not necessarily true in strict analysis. 
Recent experiments \cite{Delft} claim to have closed all ``closable" loopholes.

There have been numerous attempts to go beyond the CHSH inequalities.
 Mermin introduced a set of inequalities for an arbitrary number of qubits that were maximally violated
 by the GHZ state \cite{Mermin,GHZ}. A systematic mathematical treatment of these inequalities was carried
 out a decade later \cite{WernerWolf,Cereceda,Zukowski}. It was also at that time that an inequality for two parties, each performing quantum measurements with
$d$ outcomes was discovered \cite{CollinsLinden} and with it came the first realization that maximally entangled states do not always maximally violate a Bell inequality \cite{Latorre}, which showed that entanglement is not in a one-to-one correspondence with nonlocality. Progress in generalization to a larger number of d-dimensional particles has been more modest \cite{Acin}. For a general recent review of Bell nonlocality and a large list of references, see Ref. \cite{Brunner}.

The main aim of this paper is to construct Bell inequalities for systems composed of several subsystems composed by more than two levels each. In particular, we focus our attention on quantum systems consisting on qutrits. Inequalities for three outcomes have been written more often in terms of probabilities but they can also be treated with expectation values \cite{Chen,Arnault}. We have extended this formalism in order to build new inequalities for three outcomes and a different number of parties and find its classical and quantum bounds for qutrits in a semi-systematic way. We have found some regular patterns for the coefficients of the inequalities and for the settings and states that maximally violate these inequalities. This mechanism is potentially generalizable to other dimensions.

This work is organized as follows. In Sect. \ref{qubits}, a review of the CHSH and Mermin inequalities for two outcomes and several parties is presented. We focus on an interesting pattern involving commutators, which we use to write n-particle inequalities and classical and quantum bounds in a simple way. 
In Sect. \ref{qutrits}, the work done for qutrits is reviewed and we present our formalism and methods to construct new inequalities and find their classical and quantum bounds. In Sect. \ref{mapping}, a new strategy is presented to find Bell inequalities from the expressions of maximally entangled states.  
Some further issues, including the multiplets of optimal settings (MOS) and 
potential generalization of the results obtained for higher dimensions are discussed in the Appendix.

\section{Bell inequalities for two outcomes}\label{qubits}

\subsection{Two parties}

In the case of two parties the only relevant Bell inequality is the one of 
Clauser, Horne, Shimony and  Holt \cite{CHSH}. It is obtained out of the following Bell polynomial
\begin{equation}
B_{CHSH} = ab + ab' +a'b -a'b'.
\label{chshn}
\end{equation}
Here, $a,a'=\pm 1$ and $b,b'=\pm 1$ are the possible outcomes detected by observers Alice and Bob, respectively. Note that Eq.(\ref{chshn}) can be factorized as
\begin{equation}
B_{CHSH} = a(b+b')+a'(b-b'),
\label{chshbrackets}
\end{equation}
so one of the terms is $\pm 2$, while the other one is equal to zero, 
which means that the maximum value that can be obtained with a local realistic theory is $\langle B_{CHSH}\rangle_{LR}=2$. In a more general case, this \textit{classical} bound can be obtained by computing the value of the Bell polynomial with all possible outcomes for $a,a',b$ and $b'$ and selecting its maximum.

In quantum mechanics, the variables $a,a'$ and $b,b'$ are represented by Hermitian operators acting on the Hilbert spaces $\mathcal{H}_{a}$ and $\mathcal{H}_{b}$, respectively. For dichotomic variables the operators satisfy $a^{2}=a'^{2}=b^{2}=b'^{2}=\mathbb{I},$ because the measurement operators $a,a',b$ and $b'$ have eigenvalues $\pm 1$. The quantum Bell operator reads then
\begin{equation}\label{CHSH2}
B_{CHSH} = a \otimes b + a \otimes b' +a' \otimes b -a' \otimes b',
\end{equation}
where $\otimes$ denotes the Kronecker product. The \emph{quantum} bound $\langle B_{CHSH}\rangle_{QM}$ corresponds to the maximal eigenvalue of all possible Bell operators (\ref{CHSH2}) satisfying the previously stated conditions. A Bell operator $B$ defines a Bell inequality if $\langle B \rangle_{LR}<\langle B\rangle_{QM}.$
In the case of CHSH, it was proven by Tsirelson \cite{Tsirelson} that the maximum quantum value is $\langle B_{CHSH}\rangle_{QM}=2\sqrt{2}$.
An enlightening proof of this quantum value is given in Ref. \cite{Landau} and is reproduced now. The square of the Bell operator shown in Eq. \eqref{CHSH2} is
\begin{equation}
B^{2}_{CHSH}= 4\mathbb{I}_a\otimes \mathbb{I}_b-[\hat{a},\hat{a'}]\otimes[\hat{b},\hat{b'}]\ .
\label{chsh2o}
\end{equation}
For a local hidden variable theory all observables commute, so the classical value is determined by $\langle B_{CHSH}\rangle_{LR}=\sqrt{\langle B^2_{CHSH}\rangle_{LR}}=\sqrt{4}=2.$
On the other hand, the largest absolute value of all the possible eigenvalues for commutators of hermitian operators is 2 and it is achieved by considering the Pauli matrices, as they have the property
$[\sigma_j,\sigma_k]=2i \epsilon_{jkl} \sigma_l$ and $\sigma_l$ has eigenvalues $\pm 1$. Here, $\epsilon_{jkl}$ is the antisymmetric Levy-Civita tensor. Therefore, the quantum value of the square Bell operator \eqref{chsh2o} is given by $
\langle B_{CHSH}\rangle_{QM}=\sqrt{\langle B^2_{CHSH}\rangle_{QM}}=\sqrt{8}=2\sqrt{2}.$
In Sect. \ref{mermin} we give a more formal treatment of this technique.

It is interesting to study the ratio associated to a Bell polynomial
\begin{equation}
R(B)=\frac{\langle B\rangle_{QM}}{\langle B\rangle_{LR}},
\end{equation}
as it quantifies the strength of the inequality generated by the Bell operator $B$.
Note that a Bell inequality is characterised by the ratio $R(B)>1$. 
For example, for the CHSH inequality we have $R(B_{CHSH})=\sqrt{2}$.

Quantum states producing $R(B)>1$ are non-local in the sense that those ratios cannot be reproduced by considering a local hidden variable theory. As consequence, non-local quantum states cannot be fully separable. However, entanglement and non-locality are different concepts. Indeed, some entangled states do not violate any Bell inequality. Furthermore, states producing the maximal ratio are typically highly entangled \cite{Kar}.

This paper focuses on the study of this ratio, although more elaborated measures can be studied, like the p-value \cite{Delft} or the Kullback-Leibler relative entropy \cite{vandam}. 

\subsection{Three parties}\label{TP}

In the case of three qubits the most general symmetric Bell operator can be written as
\begin{eqnarray}\label{B3}
B_{3}&=& z_{0} (a\otimes b\otimes c)+z_{3} (a'\otimes b'\otimes c')+\nonumber \\ &&z_{1} (a\otimes b\otimes c'+a\otimes b'\otimes c+a'\otimes b\otimes c)+ \nonumber \\
 &&z_{2} (a\otimes b'\otimes c'+a'\otimes b\otimes c'+a'\otimes b'\otimes c),
\end{eqnarray}
where $z_0,\dots,z_3\in\mathbb{R}$. The following values for $z_i$ \cite{Moradi}
\begin{equation}
z_{i}^M=\lbrace z_0,z_1,z_2,z_3 \rbrace^M = \lbrace 0,1,0,-1 \rbrace,
\end{equation}
lead us to the 3-qubit Mermin operator
\begin{eqnarray}
M_{3}&=& (a\otimes b\otimes c'+a\otimes b'\otimes c+a'\otimes b\otimes c)-\nonumber\\ &&(a'\otimes b'\otimes c'),
\label{mermin3}
\end{eqnarray}
having a square
\begin{eqnarray}\label{eq:C3a}
M_{3}^{2}&=&4\mathbb{I}_{ABC}-\bigl([a,a']\otimes [b,b']\otimes\mathbb{I}_C+\\
&&[a,a']\otimes\mathbb{I}_B\otimes [c,c']+\mathbb{I}_A\otimes [b,b']\otimes [c,c']\bigr).\nonumber
\end{eqnarray}
For brevity the symbols of the Kronecker product and identities are suppressed in every subsequent equation.
Eq.(\ref{eq:C3a}) allows us to obtain the classical value $\langle M_3\rangle_{LR}=2$ and the quantum value $\langle M_3\rangle_{QM}=4$, since each commutator can achieve a maximum absolute value of 2.

A different set of coefficients $z_{i}^S=\lbrace 1,1,-1,1\rbrace$ was proposed by Svetlichny \cite{Svetlichny}. This choice leads to the form
\begin{eqnarray}\label{S_3}
S_{3}&=& (abc)+ (abc'+ab'c+a'bc) \nonumber \\
 && -(ab'c'+a'bc'+a'b'c) - (a'b'c'),
\end{eqnarray}
having the square form
\begin{eqnarray}
S_{3}^{2}&=&8-2\left([a,a'][b,b']+[a,a'][c,c']+[b,b'][c,c']\right)-\nonumber\\
&&\lbrace a,a'\rbrace \lbrace b,b'\rbrace \lbrace c,c'\rbrace.
\label{eq:C3b}
\end{eqnarray}
Note that this squared operator includes both commutators and anticommutators. For Pauli matrices $\lbrace\sigma_{i},\sigma_{j}\rbrace=2\delta_{ij}$, so a maximal value for the commutator implies a minimum value for the anticommutator, and vice versa. The commutators vanish
while estimating the classical value and $\langle S_3\rangle_{LR}=4$. For the quantum value the optimal case occurs when the commutators take the maximum amplitude $\pm2$ and the anticommutators vanish, so that $\langle S_{3}\rangle_{QM}=4\sqrt{2}$. The ratios for the Bell operators of Eqs. (\ref{mermin3}) and (\ref{S_3}) are given by
$R(M_3)=2\,\mbox{ and }\, R(S_3)=\sqrt{2}.$
It is known that Mermin inequality generated by the Bell operator (\ref{mermin3}) can be violated by biseparable states, whereas Svetlichny inequality defined by the operator (\ref{S_3}) cannot. Bell inequalities generated by operators like $S_3$ are called \emph{multipartite Bell inequalities}. This topic is analysed in detail by Collins et al. \cite{Collins2}.

These inequalities are already well tested experimentally. Violation of inequalities  $M_3$ and $S_3$ have been reported in Ref. \cite{mermin-3qubits} and \cite{svetexp}, respectively.

\subsection{Mermin polynomials}
\label{mermin}

There exists an entire family of $n$-qubit inequalities first discovered by Mermin \cite{Mermin,WernerWolf}. Here, we construct Mermin operators as in Ref. \cite{Collins2}. Let us change the notation of observables $\{a,b,c...\} \equiv \{a_1,a_2,a_3...\}$, which is more convenient to treat the multipartite case. Defining $M_1 \equiv a_1$, the Mermin polynomials are obtained recursively as
\begin{equation}
M_n=\frac{1}{2} M_{n-1}(a_n+a'_n) + \frac{1}{2}M'_{n-1}(a_n-a'_n),
\label{generalmermin}
\end{equation}
where $M'_k$ is obtained from $M_k$ by interchanging primed and nonprimed observables $a_n$. In particular, $M_2$ and $M_3$ correspond to the operators \eqref{CHSH2} and \eqref{mermin3}, respectively, up to a constant factor. It was proven in \cite{Cereceda} that all Mermin operators have a square form composed by  the identity and commutators, as operators \eqref{CHSH2} and \eqref{mermin3}. Let us now proceed with our version of the proof. The square of Mermin operators gives an expression containing commutators $[\cdot,\cdot]$ and anticommutators $\{\cdot,\cdot\}$
\begin{eqnarray}
M^2_n  = &&\frac{1}{4} \bigl(M^2_{n-1}(2+\{a_n,a'_n\})+M'^2_{n-1}(2-\{a_n,a'_n\}) \nonumber \\ 
&&-[M_{n-1},M'_{n-1}][a_n,a'_n]\bigr), 
\label{msquaredrec}
\end{eqnarray}
\begin{eqnarray}
M'^2_n = &&\frac{1}{4} \bigl(M'^2_{n-1}(2+\{a_n,a'_n\})+M^2_{n-1}(2-\{a_n,a'_n\}) \nonumber \\
&&-[M_{n-1},M'_{n-1}][a_n,a'_n]\bigr).
\end{eqnarray}
Furthermore, if $M^2_{n-1}=M'^2_{n-1}$, then $M^2_{n}=M'^2_{n}$. As this is true for $M^2_1=M'^2_1=1$, by induction it is true for every $n$. Therefore, Eq.\eqref{msquaredrec} can be simplified to
\begin{equation}
M^2_n = M^2_{n-1} - \frac{1}{4} [M_{n-1},M'_{n-1}] [a_n,a'_n],
\end{equation}
where
\begin{equation*}
[M_{n-1},M'_{n-1}] = [M_{n-2},M'_{n-2}] + M^2_{n-2} [a_{n-1},a'_{n-1}].
\end{equation*}
Given that $[M_1,M'_1]=[a_1,a'_1]$ every operator $M^2_n$ can be expressed as a sum of products of an even number of commutators. Thus the operator $M^2_n$ reads,
\begin{equation}
M^2_n = 1 + \sum_{s=1}^{[n/2]} \frac{(-1)^s}{2^{2s}} \sum_{i_j \in D} \prod_{j=1}^{2s} [a_{i_j},a'_{i_j}],
\label{msquared}
\end{equation}
where $D$ is the set of $n$ operators taken in groups of $2s$ elements. This result is implicitly presented in Ref. \cite{WernerWolf}. The classical and quantum values arise immediately. On one hand, $\langle M_n\rangle_{LR}=1$, as the second term in Eq. \eqref{msquared} is always zero due to the presence of commutators. On the other hand, for the quantum value every commutator takes $\pm2$, conveniently chosen to maximize it. Thus,
\begin{equation}
\langle M^2_n\rangle_{QM} = 1 + {{n}\choose{2}} + {{n}\choose{4}}+... = 2^{n-1}.
\label{msquaredquant}
\end{equation}
The quantum value for $M_n$ is, therefore, $\langle M_n\rangle_{QM}=\sqrt{\langle M^2_n\rangle_{QM}}=2^{\frac{n-1}{2}}$,
which matches the rate computed by Werner and Wolf \cite{WernerWolf}.
Let us note that when computing this last step it is assumed that the maximum eigenvalue of a sum of matrices is equal to the sum of the maximum eigenvalues, a fact that is not true in general but is true in this case.

The optimal states for the Mermin inequalities are the GHZ states \cite{Mermin,WernerWolf}. 
For $n=2$ and $n=3$ these states can be considered as maximally entangled.
However, for $n\ge 4$ it is not the case \cite{HS00,Sc04}
if one considers the mean entropy of a reduced density matrix, averaged over
all possible choices of $[n/2]$ subsystems, which define the reduced state. Here, $[x]$ denotes the integer part of $x$. Therefore, the Mermin inequalities 
provide an example, for which the maximal violation does not correspond to maximally entangled states. Let us mention that the experimental violation of Mermin inequalities has been verified up to 14 qubits with ion traps \cite{merminmany}. Recently, the $M_3, M_4$ and $M_5$ cases have been implemented on a 5 superconducting qubits quantum computer designed by IBM \cite{AL16}.

\section{Bell inequalities for three outcomes} 
\label{qutrits}
In this section we study Bell inequalities for three outcomes and their maximum violations in the case of hermitian and unitary setting operators. We remark that all the maximal violations presented for Bell inequalities and having three outcomes have been found for qutrit states. Therefore, they are lower bounds for the maximal possible quantum value which, in principle, could be attained for qudits with more than three number of levels each.

\subsection{Two parties with hermitian operators}
A Bell inequality for two parties, two settings and $d$ outcomes was proposed by Collins et al. \cite{CollinsLinden} and it is known as CGLMP inequality. The violation of some of these inequalities has been verified experimentally \cite{qutritsexp}. In the case of three outcomes the inequality is given by
\begin{eqnarray}
&p(a=b)+p(b=a'+1)+p(a'=b')+& \nonumber \\
&p(b'=a)-p(a=b-1)-p(b=a')-\nonumber \\
&p(a'=b'-1)-p(b'=a-1) \leq 2,&
\label{CGLMP}
\end{eqnarray}
where the possible outcomes are $\{0,1,2\}$ and the sum inside probabilities is modulo $d=3$. This Bell inequality can be associated with the following Bell operator 
\begin{eqnarray}
C_{223}&=&2-3(a^2+b'^2)+\frac{3}{4}(ab+a^2b-a'b-a'^2b-ab^2+\nonumber\\
&&a'b^2+ab'-a^2b'+a'b'+a'^2b'+ab'^2-a'b'^2)+\nonumber\\
&&\frac{9}{4}(a^2b^2-a'^2b^2+a^2b'^2+a'^2b'^2),
\label{cglmpreal}
\end{eqnarray}
where the notation $C_{nsd}$ stands for $n$ parties, $s$ settings and $d$ outcomes. The optimal settings can be obtained by choosing one arbitrary setting and obtaining the other one with a phase transformation followed by the Fourier transform, as discussed extensively in Ref. \cite{CollinsLinden}. The quantum value is given by $\langle C_{223}\rangle_{QM}=2(5-\gamma^2)/3\approx2.9149$ for the optimal state $|\psi\rangle= (|00\rangle+\gamma|11\rangle+|22\rangle)/\sqrt{(2+\gamma^2)}$ where $\gamma =(\sqrt{11}-\sqrt{3})/2 \approx 0.7923$ \cite{Latorre}. The violation rate for this quasi Bell state reads $R_{2t}=(5-\gamma^2)/3 \approx 1.4547$. In Ref. \cite{Latorre} the ratios for CGLMP inequalities are found up to $d=8$ levels. The optimal settings can be conveniently expressed in terms of eight Gell-Mann matrices $\lambda_i$, 
the traceless generators of SU(3) \cite{G64}.
The optimal settings for the Bell inequality generated by the operator (\ref{cglmpreal}) are
\begin{eqnarray}
A=B&=&\lambda_3, \nonumber \\
A'=B'&=&\frac{2}{3} (\lambda_1+\lambda_6) + \frac{1}{6} (\lambda_3+\sqrt{3}\lambda_8).
\label{gellmann2qt}
\end{eqnarray}
where $J_1$ and $J_3$ are two elements of the representation of SU(2) in three dimensions. 

The Bell operator in Eq. \eqref{cglmpreal} has a rather long and unenlightening form. In the next subsection we will show how the consideration of unitary setting operators instead of hermitian operators simplifies the study of Bell inequalities.

\subsection{Two parties with unitary operators}

A more convenient way to represent Bell inequalities for three outcomes is by considering complex outcomes associated to the third roots of unity \cite{Zukowski2,Chen,Arnault}. In this way, settings turn from hermitian to unitary operators with eigenvalues $\{1,w,w^2\}$, where $w=\exp(2\pi i/ 3)$. Note that for qubits the Pauli matrices are both hermitian and unitary, while for qutrits a choice between one of these properties has to be made. Note that any operator that can be expressed as a linear combination 
(with real or complex coefficients) of rank one projectors forming a POVM allows for a physical interpretation. Note also that sum of unitary operators is in general, not a normal operator. 
A complex operator $M$ is normal if $[M,M^{\dag}]=0$. However, any operator 
can be decomposed into its \emph{hermitian} 
and \emph{anti-hermitian} part, $B=[B]_H +i [B]_A$, where
$[B]_H:= \frac{1}{2}(B+B^\dagger)$ and
$[B]_A:= \frac{1}{2i}(B-B^\dagger)$ are hermitian operators and, therefore, they have real eigenvalues.

The Bell operator \eqref{cglmpreal} can be written as
the anti-hermitian part of a non-hermitian operator,
\begin{equation}
C_{223}=  \left[ a(w b-b')+a'(wb'-b) \right]_A.
\label{cglmpim}
\end{equation}
This form appears to be a direct generalization of the CHSH operator \eqref{chshbrackets}, with different signs and relative phases added. If one of the terms reaches the maximum value $\sqrt{3}$ then the other one is forced to be zero. The classical and quantum values for this operator are $\langle C_{223}\rangle_{LR} = \sqrt{3}\approx 1.73$ and $\langle C_{223}\rangle_{QM}  = (1/2)(\sqrt{3}+\sqrt{11})\approx 2.52$, and the ratio is given by $R(C_{223}) = (1/3)(5-\gamma^2)\approx 1.45$. The violation rate is therefore the same as for CGLMP inequality \eqref{CGLMP} as expected, because it is the same inequality albeit written in a different language. Let us now find the optimal settings for the operator (\ref{cglmpim}). The convenient representation for unitary operators are the generalized unitary Pauli matrices which form the Weyl-Heisenberg group. The generators of the group are
\begin{equation}\label{XZ}
X = \left( \begin{array}{ccc}
0 & 0 & 1 \\
1 & 0 & 0 \\
0 & 1 & 0 \end{array} \right)\hspace{0.5cm}\mbox{and}\hspace{0.5cm}
Z = \left( \begin{array}{ccc}
1 & 0 & 0 \\
0 & w & 0 \\
0 & 0 & w^2 \end{array} \right),
\end{equation}
where $\omega=e^{2\pi i/3}$. An orthonormal basis is given by the nine elements
\begin{equation}
X^k Z^j = \sum_{m=0}^2 |m+k\rangle w^{jm} \langle m|\, ,
\label{genpauli}
\end{equation}
which are proportional to the elements of the Weyl-Heisenberg group. By numerical optimization it is possible to show that the optimal settings for the operators \eqref{cglmpim} are
\begin{eqnarray}
&A=B=X,&\nonumber \\
&A'=B'=\frac{1}{3} (- X + 2w XZ + 2w^2 XZ^2).&
\label{complexap}
\end{eqnarray}
In matrix notation, $A'$ has a simple structure
\[ A' = \left( \begin{array}{ccc}
0 & 0 & 1 \\
-1 & 0 & 0 \\
0 & -1 & 0 \end{array} \right).\] 
The optimal settings for all the complex CGLMP inequalities, in this case ($\{X,A'\}$), are called \emph{multiplets of optimal settings} (MOS). In Appendix \ref{mubs} some properties of MOS are discussed.

Let us investigate the square of the operator $C_{223}$ introduced in
(\ref{cglmpreal}). Making use of the identity for the hermitian and antihermitian
 parts of an operator $C$
\begin{equation}
 (C_A)^2 = \frac{1}{4}(CC^\dagger + C^\dagger C) - \frac{1}{2} (C^2)_H \, ,
\label{cglmpim2}
\end{equation}
it is easy to show that $C_{223}C_{223}^\dagger$ has an interesting structure
\begin{equation}
C_{223}C_{223}^\dagger = 3 + (1+\{\{a,a'\}\})(1+\{\{b,b'\}\}) .
\label{ccdagger}
\end{equation}
Here  $\{\{a,a'\}\}$ is called the \emph{complex anticommutator} $\{\{a,a'\}\}= aa'^\dagger + a'a^\dagger$. The complex anticommutator attains its maximum value 2 both for MOS and MUB (see appendix \ref{mubs} for a definition of these pairs of matrices). However, its classical value can also be equal to 2 by using $a=a'=1$. Thus the form (\ref{ccdagger}) does not allow us to distinguish between classical and quantum values.

\subsection{Three parties}

A three parties Bell inequality was proposed by Ac\'in et al. in Ref. \cite{Acin}. 
In the probability formalism it reads 
\begin{eqnarray}
&p(a+b+c=0) + p(a+b'+c'=1) +&\nonumber \\
&p(a'+b+c'=1)+p(a'+b'+c=1)-&\nonumber \\
&2p(a'+b'+c'=0) - p(a'+b+c=2) -& \nonumber \\
& p(a+b'+c=2)- p(a+b+c'=2) \leq 3.&
\label{3qtprob}
\end{eqnarray}
The analysis here is very similar to the CGLMP case: the maximal violation is given by a quasi maximally entangled state $|\psi\rangle= (|000\rangle+\gamma |111\rangle+|222\rangle)/\sqrt{2+\gamma^2}$ where now $\gamma\approx1.186$. The quantum value is $4.37$ and the violation rate is $R=(5-\gamma^2)/3 \approx 1.4574$, as for 2 qutrits. The corresponding hermitian Bell operator has a rather long form, so we will not reproduce it here. The optimal settings can be expressed in terms of the Gell-Mann matrices as
\begin{eqnarray}
&&A=B=C=\lambda_3,\nonumber \\
&&A'=B'=C'=\frac{1}{\sqrt{3}} (\lambda_2+\lambda_4+\lambda_6) \, .
\end{eqnarray}
Let us now consider the case of unitary settings having complex eigenvalues. The Bell operator associated to inequality \eqref{3qtprob} can be expressed as hermitian part of an operator
\begin{eqnarray}
C_{333}&=&\mathbb{I}+\frac{2}{3}  \bigl[abc +2a'b'c' +w(a'b'c+a'b c'+ab'c') \nonumber \\
&&-w^2(a'b c+a b'c +a b c')\bigr]_H.
\end{eqnarray}
One can also drop the additive and multiplicative terms and study the simplified operator
\begin{eqnarray}
C'_{333}&=&  \bigl[abc +2a'b'c' +w(a'b'c+a'b c'+ab'c') \nonumber \\
&&-w^2(a'b c+a b'c +a b c')\bigr]_H.
\label{3qutrits}
\end{eqnarray}
Here, the classical value is $\langle C'_{333}\rangle_{LR}=3$ and the quantum value is $\langle C'_{333}\rangle_{QM}=(3/4)(1+\sqrt{33})\approx5.058$, which yields to the ratio $R(C'_{333}) = (1/4)(1+\sqrt{33})\approx1.686$. The optimal settings are given by
\begin{eqnarray}
&A=B=C=X,& \nonumber \\
&A'=B'=C'=Z.&
\end{eqnarray}
Note that the settings are mutually unbiased (see Appendix \ref{mubs}). Now the violation rate is greater because the additive constant term has been eliminated. This appears somewhat arbitrary but it is more convenient to compare  inequalities  for two and three qutrits without additive terms. In this way, it is expected that the rate of violation increases with the number of particles, as it happens for qubits. Intriguingly, the 3-qutrit operator \eqref{3qutrits} can be derived from the 2-qutrit CGLMP operator \eqref{cglmpim} and adding a third party such that the resulting 3-qutrit operator is symmetric, as shown in Appendix \ref{2to3}.

\subsection{Larger number of parties}
In the case of four parties, two settings and three outcomes we have found the following symmetric Bell operator
%
\begin{eqnarray}
&&C_{423}=  \bigl[2(abcd) + (a'bcd+ab'cd+abc'd+abcd')   \nonumber \\
&&+w(a'b'cd+a'bc'd+a'bcd'+ab'c'd+ab'cd'+abc'd') \nonumber \\
&&+(a'b'c'd+a'bc'd'+a'b'cd'+ab'c'd') +2 (a'b'c'd')\bigl]_A,\nonumber \\
\label{4qt}
\end{eqnarray}
which produces $\langle C_{423}\rangle_{LR}=3\sqrt{3}\approx5.19$, $\langle C_{423}\rangle_{QM}\approx 9.766$ and $R(C_{423})\approx 1.879$ for the optimal settings
\begin{eqnarray}
&&A=B=C=D=X, \nonumber \\
&&A'=B'=C'=D'=Z,
\end{eqnarray}
which are again mutually unbiased settings. The optimal state has entanglement properties equivalent to those of the GHZ of four parties and three settings $|GHZ_{4,3}\rangle=(|0000\rangle+|1111\rangle+|2222\rangle)\sqrt{3}$.

For 6 parties we have also found a symmetric Bell operator. To simplify the notation, the polynomials having terms with the same number of primes are denoted by its number of primes in parenthesis, for example:
$(1') \equiv a'bcdef+ab'cdef+abc'def+abcd'ef+abcde'f+abcdef'$. In this notation, the 6 parties operator reads
\begin{equation}
C_{623}=-w(0')+(1')-(2')+w(3')-(4')+(5')-w(6').
\label{primenotation}
\end{equation}

For this inequality, $\langle C_{623}\rangle_{LR}=9\sqrt{3}\approx15.589$, $\langle C_{623}\rangle_{QM}\approx32.817$ and $R(C_{623})\approx2.105$, with MOS optimal settings. The maximal violation is a given by a \emph{quasi} GHZ state, as for the case of 2 and 3 qutrits. \\

Let us summarize the results for the symmetric Bell operators for $n$-qutrit systems studied in this section. Unfortunately, we could not find a 5-qutrit inequality that follows all the patterns. The inequalities considered are those determined by the coefficients of Table \ref{tabcoefficients}, and the results are summarized in Table \ref{tabresults}.

\begin{table}[h!]
\centering
\begin{tabular}{c | c | c | c | c | c}
\hline
\backslashbox{Terms}{Parties} & 2 & 3 & 4 & 5  & 6 \\
\hline
(0') & $\omega$ & $1$    & $2$   & $\omega^2$  & $-\omega$ \\
(1') & $1$ & $-\omega^2$ & $1$   & $-\omega^2$ & $1$ \\
(2') & $\omega$ & $\omega$    & $\omega$   & $-\omega^2$ & $-1$ \\
(3') &     & $2$    & $1$   & $-\omega^2$ & $\omega$ \\ 
(4') &     &        & $2$   & $\omega^2$  & $-1$ \\  
(5') &     &        &       & $\omega^2$  & $1$ \\
(6') &     &        &       &        & $-\omega$ \\
\hline
\end{tabular}
\caption{Coefficients for symmetric Bell inequalities  from two to six parties and three settings and three outcomes, where $\omega=e^{2\pi i/3}$. The primed notation $(k')$ identifies all terms having $k$ primed settings, as defined before in Eq.\eqref{primenotation}.}
\label{tabcoefficients}
\end{table}

\begin{table}[h!]
\begin{tabular}{c | c | c | c | c | c}
\hline
\backslashbox{}{\hspace{2mm}Qutrits} & 2 & 3 & 4 & 5 & 6 \\
 \hline
$\langle [B]_A\rangle_{LR}$ & \boldmath{$\sqrt{3}$} & $3\sqrt{3}$ & \boldmath{$3\sqrt{3}$} & $9\sqrt{3}$&  \boldmath{$9\sqrt{3}$}\\
$\langle [B]_A\rangle_{LR}^{(-)}$ & $-2\sqrt{3}$ & $-3\sqrt{3}$ & $-6\sqrt{3}$ & $-9\sqrt{3}$ & $-18\sqrt{3}$\\
$\langle [B]_H\rangle_{LR}$ & $3$ & \boldmath{$3$} & $9$ & \boldmath{$9$} & $27$\\
$\langle [B]_H\rangle_{LR}^{(-)}$ & $-3$ & $-6$ & $-9$ & $-18$ & $-27$ \\
$\langle [B]_x\rangle_{QM}$ & $ 2.524 $ & $5.058$ & $9.766$ & $\mathit{15.575}$ & $32.817$ \\
$\mathrm{R}$ & $1.457$ & $1.686$ & $1.879$ & $\mathit{1.731}$ & $2.105$ \\
$Settings$ & MOS & MUB & MUB & \it{Num.} & MOS \\
P & $0.347$ & $0.342$ & $1/3$ & $\mathit{0.351}$ & $0.334$\\ 
\hline
\end{tabular}
\caption{Main results for inequalities from 2 to 6 qutrits, where it can be seen that the classical patterns match perfectly, while the 5-qutrit inequality appears not to follow the quantum pattern. Here, $\langle B\rangle_{LR}$ and $\langle B\rangle_{LR}^{(-)}$ denote the maximum and minimum classical value for optimizations of anti-hermitian or hermitian part of the operator, respectively. The quantity that we take as the extremal classical bound is marked in bold, and
$\langle [B]_x\rangle_{QM}$ stands for its corresponding quantum value, where $x=A$ for an even number of qutrits and $x=H$ for an odd number of qutrits. $R=\langle B\rangle_{QM}/\langle B\rangle_{LR}$ and \emph{Settings} denotes the optimal settings. $P$ denotes the purity of the $\lfloor n/2 \rfloor$ party reductions of the optimal state and \emph{Num.} means numerical approximate solution, and italic font in the 5-qutrits case is written to note that this case does not follow the same patterns of the others. We remark that optimal values appearing in this table have been achieved by optimizing over qutrit systems.}
\label{tabresults}
\end{table}

The main patterns that can be seen in Table \ref{tabresults} are
\begin{itemize}
\item[\emph{(i)}] 
For an even number of qutrits 
the classical values $\langle B\rangle_{LR}$ arise from the anti-hermitian part of an operator 
while for odd number of qutrits one takes its hermitian part.
The following relation between the minimal and the maximal classical values
holds, $\langle B^-\rangle_{LR}=-2\langle B\rangle_{LR}$.

\item[\emph{(ii)}] There is a factor of $\sqrt{3}$ between the maximum value of the 
hermitian and anti-hermitian parts, 
and also a factor of $\sqrt{3}$ between the maximal value of two consecutive numbers of qutrits. The maximal value of the hermitian parts are the same for $n$ and $n+1$ qutrits if $n$ is even. Also, the maximal value of the anti-hermitian parts are the same for $n$ and $n+1$ if $n$ is odd.

\item[\emph{(iii)}] The quantum value $\langle B\rangle_{QM}$ of a non-hermitian operator $B$ is computed as the maximum over quantum values of the hermitian and anti-hermitian parts, i.e.,  $\langle B\rangle_{QM}=\mathrm{Max}\{\langle B_H\rangle_{QM},\langle B_A\rangle_{QM}\}$. The rate of violation increases with the number of qutrits except for the 5-qutrit case, which do not follow the patterns.

\item[\emph{(iv)}] The optimal settings are either MUB or MOS, with the exception of the 5-qutrit inequality.

\item[\emph{(v)}] The optimal states have entanglement properties close to a GHZ or exactly those of a GHZ state in the case of four qutrits. In Table \ref{tabresults} the closeness to the GHZ state is measured by the purity $P$ of the reduced matrix $\sigma$ over $\lfloor n/2 \rfloor$ particles. The GHZ state of $n$ qutrits has reductions to two parties with $P={\rm Tr} \sigma^2=1/3$, 
whereas the absolutely maximally entangled state has $P=1/3^{[n/2]}$ 

\end{itemize}

\section{Mapping states to Bell operators}\label{mapping}

Let us now present a novel idea to generate Bell inequalities based on a mapping from maximally entangled states to Bell operators. We shall illustrate the construction through an example and, then, generalize it to different cases. 

The two-qubit state
\begin{equation}
|\psi \rangle = (|+\rangle\otimes\,|0\rangle +|-\rangle\otimes\,|1\rangle)/\sqrt{2},
\end{equation}
where $|\pm\rangle=\sqrt{1/2}\bigl(|0\rangle\pm|1\rangle\bigr)$, can be expanded to match the form
\begin{equation}\label{mes}
| \psi \rangle = \frac{1}{2} \left( |0_A0_B\rangle + |0_A1_B\rangle + |1_A0_B\rangle - |1_A1_B\rangle \right).
\end{equation}
This state belongs to the set of maximally entangled  Bell states.
The CHSH Bell operator can be obtained from this state by identifying first and second particle with observables for Alice and Bob, respectively. We identify symbol $0$ with non-primed settings and symbol $1$ with primed settings, as in Table \ref{legend}.
\begin{table}[h!]
\centering
\begin{tabular}{c  c  c}

   $| \psi \rangle$ & $\rightarrow$ & $\mathcal{B}$ \\
\hline
   $| 0_A \rangle$ & $\rightarrow$ & $a$ \\
   
   $| 1_A \rangle$ & $\rightarrow$ & $a'$ \\
   
   $| 0_B \rangle$ & $\rightarrow$ & $b$ \\
   
   $| 1_B \rangle$ & $\rightarrow$ & $b'$ \\

\end{tabular}
\caption{Substitution legend for mapping states to Bell operators for the CHSH case.}
\label{legend}
\end{table}

By removing the normalization term the CHSH operator arises
\begin{equation}
\mathcal{B}_{CHSH} = ab + ab' + a'b - a'b' \, .
\end{equation}
Furthermore, the maximally entangled state (\ref{mes}) is the optimal state for a suitable choice of the measurement settings. This remarkable fact motivates us to study new multipartite Bell inequalities generated from multipartite quantum states.

\subsection{Bell inequalities from entangled states}\label{BIFES}

The general strategy is to construct Bell inequalities associated 
to some distinguished maximally entangled states.
Starting from the Bell state for two qutrits,
$|\psi_3^+\rangle=(|00\rangle+|11\rangle+|22\rangle)/\sqrt{3}$
 and applying the Fourier transform to the second party we obtain 
\begin{equation}\label{IF_GHZ}
|\phi\rangle=\mathbb{I}\otimes F_3|\psi_3^+\rangle.
\end{equation}
From this state, using legend from table \ref{legend} and adding the case $|2_A \rangle \rightarrow a''$ and analogously for party B, a new Bell operator for 2 qutrits and 3 settings arises,
\begin{equation}
C_{233} = [\,\vec{a}\cdot F_3\vec{b}\,]_H,
\end{equation}
where $\vec{a}=(a,a',a'')$, $\vec{b}=(b,b',b'')$ and $F_3$ is the Fourier matrix of order three, $(F_3)_{jk}=e^{2\pi ijk/3}$. This operator has a classical value $\langle C_{233}\rangle_{LR}=9/2$ and it is maximally violated by a state with the same entanglement properties of the GHZ with a violation ratio $R(C_{233})=2/\sqrt{3} \cos (\pi/18)\approx1.137$ for the optimal MUB settings
\begin{eqnarray}
&&A=B=X , \nonumber \\
&&A'=B'=Z , \nonumber \\
&&A''=B''=X^2 Z^2 ,
\end{eqnarray}
where $X$ and $Z$ are given in Eq.(\ref{XZ}). An equivalent inequality with the same properties was found in Refs. \cite{koreans1, koreans2}.

We can apply the same strategy for four qutrits 
starting with the GHZ state 
$|GHZ_4^3\rangle=(|0000\rangle + |1111\rangle +|2222\rangle)/\sqrt{3}$. 
Acting with Fourier transform $F_3$ on three parties we obtain a locally
equivalent state
\begin{equation}\label{IFFF_GHZ}
|GHZ_4^{3'}\rangle=\mathbb{I}\otimes F_3\otimes F_3\otimes F_3|GHZ_4^3\rangle,
\end{equation}
which leads to the Bell operator
\begin{equation}
C'_{433} = [\,\vec{a}\cdot F_3\vec{b}\cdot F_3\vec{c}\cdot F_3\vec{d}\,]_H,
\end{equation}
where $\vec{a}=(a,a',a'')$, $\vec{b}=(b,b',b'')$, and analogously for other parties. The generalized inner product of four vectors is defined as $w\cdot x\cdot y\cdot z=\sum_{j=0}^{2} w_j x_j y_j z_j$. The optimal state has the entanglement properties of the GHZ,
but with a larger violation ratio than for the operator \eqref{4qt}. 

\subsection{Bell inequalities from AME state}

An absolutely maximally entangled state (AME) of $n$ particles is a state with every reduction, up to $\lfloor n/2 \rfloor$ particles, maximally mixed  \cite{Goyeneche,HCLRL12,GKL15,Helwig}. Let us now try the strategy above described for the AME of 4 qutrits
\begin{equation}\label{AME43}
AME(4,3)=\frac{1}{9}\sum_{i,j,k,l=0}^2 w^{j(i-k)+l(i+k)}|ijkl\rangle.
\end{equation}

The recipe to construct the Bell operator consists in taking representation (\ref{AME43})
which contains $3^4=81$ terms with coefficients of the form $\{1,w,w^2\}$.
In the next step one uses the same legend from previous subsection. This procedure 
leads us to a Bell operator for four parties, three settings and three outcomes, 
which can be written in a compact way as
\begin{equation}
C_{433} = \sum_{i,j,k,l=0}^2 w^{j(i-k)+l(i+k)}a_{i}b_{j}c_{k}d_{l},
\label{c433}
\end{equation}
where $a_0 = a, a_1 = a', a_2 = a''$, and the same for the rest of the observables.

After transformations $d' \rightarrow wd'$ and $d' \leftrightarrow d''$, 
numerical optimization produces the following configuration of optimal settings 
\begin{eqnarray}
&&A=B=C=D=X \, , \nonumber \\
&&A'=C'=D'=X^2Z^2 \hspace{5mm} B'=X, \nonumber \\
&&A''=C''=D''=Z \hspace{9mm} B''=N,
\end{eqnarray}
where $N$ is certain matrix of size three obtained numerically. 
The optimal settings are not symmetric because the AME state 
is not symmetric under interchange of particles.

Numerical optimization suggests that the optimal state is not AME. 
Surprisingly, it has almost the same entanglement properties as the GHZ state, namely its purity is $P=1/3$ for the density matrices of reductions to 2 parties, and $P=1/3$ for three of the possible reductions to one party, while the fourth one (party B) has $P=1$, indicating that party B is in a product state with the other three. The same violation ratio as for four qutrits with two settings is obtained, see Eq. \eqref{4qt}. 
This result, and the fact that the optimal settings include $B=B'$ 
suggests that the third setting is not adding anything new 
and that this inequality is essentially the same as in the case of two settings.

Table \ref{tab3settings} summarizes the results for the 3-settings qutrit inequalities arising 
from entangled states.

\begin{table}[h!]
\begin{tabular}{c | c | c | c }
\hline
\backslashbox{}{\hspace{2mm}Qutrits} & 2 & 4 (GHZ) & 4 (AME)\\
\hline
$\langle [B]_A\rangle_{LR}$ & $3\sqrt{3}$ & $9\sqrt{3}$ & $9\sqrt{3}$ \\
$\langle [B]_A\rangle_{LR}^{(-)}$ & $-3\sqrt{3}$ & $-9\sqrt{3}$ & $-9\sqrt{3}$\\
$\langle [B]_H\rangle_{LR}$ & \boldmath{$4.5$} & \boldmath{$13.5$} & \boldmath{$13.5$} \\
$\langle [B]_H\rangle_{LR}^{(-)}$ & $-4.5$ & $-27$ & $-27$ \\
$\langle [B]_H\rangle_{QM}$ & $ 5.117 $ & $26.025$ & $25.372$ \\
R & $1.137$ & $1.928$ & $1.879$\\
$Settings$ & MUB & Num. & MUB and Num. \\
P & 1/3 & 1/3 & 1/3 \\
\hline
\end{tabular}
\caption{Characterization of Bell inequalities for 2 and 4 parties, 3 settings and 3 outcomes. There is one 4-qutrit inequality built from the GHZ state and another one built from the AME state. For all the cases the optimal states are states with the same entanglement properties as the GHZ. Abbreviations and symbols are considered as in Table \ref{tabresults}, although in this case the quantum bound is computed always with the hermitian part.}
\label{tab3settings}
\end{table}

\section{Concluding remarks}

We have used the formalism of unitary matrices with complex roots of unity as eigenvalues to construct Bell inequalities of multipartite systems, 3 settings and 3 outcomes (see Section \ref{qutrits}). We have shown that the 2-party and 3-party inequalities from Ref. \cite{CollinsLinden} and Ref. \cite{Acin} are closely related. Furthermore, we have extended these cases to 4 and 6 parties and, less convincingly, to 5 parties. We obtained regular patterns for this set of inequalities, as shown in Table \ref{tabresults}. Two of the most striking patterns are: \emph{a}) the structure of the classical bounds and their simple arithmetic progression with the number of particles, and \emph{b}) the fact that the inequalities tend to have a maximal quantum bound for settings that are either MUBs or multiplets of optimal settings (MOS) -- see Appendix \ref{mubs}.

We also introduced a mapping from entangled states to Bell operators that allows us to define Bell inequalities for multipartite systems (see Section \ref{BIFES}). In particular, we have constructed new inequalities for two and four parties with three settings, which are maximally violated by states with the same entanglement properties as the GHZ state.  We also demonstrated that a Bell inequality generated by a given quantum state is not necessarily maximally violated by the same state. For example, the inequality Eq.(\ref{c433}) is generated by the absolutely maximally entangled state of 4 qutrits, but maximally violated by a GHZ like state. This novel formalism has the potential to generate a wide range of Bell inequalities for an arbitrary large number of parties, settings and outcomes.

Let us also mention here some important questions, which remain open. Concerning the approach to Bell inequalities from squares of operators represented by commutators it would be interesting to find a procedure to determine whether a given Bell operator allows such a form. Analyzing the mapping between states and Bell operators one can raise the question whether a maximally entangled state is necessary to produce a tight Bell inequality in the case of 2 outcomes (e.g. it holds for the CHSH and all Mermin inequalities). On the other hand, the mathematical characterization of the entire set of MOS for the CGLMP inequalities defined in Appendix \ref{mubs} is a pending task. Finally, it would be interesting to have a generating polynomial for Bell inequalities with 3 outcomes, in the same way that we have the Mermin polynomials for Bell inequalities with 2 outcomes (see Eq \ref{generalmermin}).

\section*{Acknowledgments}
It is a pleasure to thank A. Cabello, T. Ac\'in and R. Augusiak for useful remarks. D.A. acknowledges financial support from the APIF Scholarship of University of Barcelona. JIL acknowledges financial support by FIS2013-41757-P. This work has been supported by the Polish National Science Center under the project number DEC-2015/18/A/ST2/00274, by the European Union FP7 project PhoQuS@UW (Grant Agreement No. 316244), by ERC grant QOLAPS (Grant No. 291348) and by the John Templeton Foundation under the project No. 56033.

\appendix

\section{Maximizing settings: mutually unbiased bases and multiplets of optimal settings} \label{mubs}

In the present work we have shown that two remarkable sets of measurement settings optimize the violation of Bell inequalities. These are the \emph{mutually unbiased bases} (MUB) and \emph{multiplets of optimal settings} (MOS). Two orthonormal bases $\{|\phi_0\rangle ,\dots ,|\phi_{d-1}\rangle \}$ and $\{|\psi_{0}\rangle ,\dots ,|\psi_{d-1}\rangle \}$ are mutually unbiased if
\begin{equation}
|\langle \phi_{j}|\psi_{k}\rangle |^{2}={\frac {1}{d}},\quad \forall j,k\in \{0,\dots ,d-1\} \, .
\end{equation}
If $d$ is a prime power number, i.e. $d=p^n$ for $p$ prime and $n\in\mathbb{N}$, then there exists a maximal set of $d+1$ MUB. In prime dimensions such set is given by the eigenvectors bases of the $d+1$ generalized Pauli operators defined in Eq.\eqref{genpauli}
\begin{equation}
X, Z, XZ, XZ^2, ..., XZ^{d-1}.
\end{equation}
We say that a set of normal operators is MUB if their eigenvectors bases are MUB.
For example, the optimal settings for Mermin inequalities for qubits are MUB. Indeed, if one setting is fixed to $\sigma_x$ then the other setting has to be a linear combination of the form $\alpha \sigma_y + \beta \sigma_z$ in order to maximize the eigenvalue of the commutator. This restriction implies that the settings are MUB.

In the qutrit case, the optimal settings for the CGLMP inequality, $A=\lambda_3$ and $A'=\frac{2}{3} (\lambda_1+\lambda_6) + \frac{1}{6} (\lambda_3+\sqrt{3}\lambda_8)$, are not MUB. However, for three qutrits the optimal settings, $A=\lambda_3$ and $A'=\frac{1}{\sqrt{3}} (\lambda_2+\lambda_4+\lambda_6)$, are MUB.

For Bell operators with complex settings the optimal settings have a more regular structure. The elements of the basis $X^i Z^j, X^k Z^l$ are MUB except for the case where $j=l$ and $i=k$. So it is clear that in 3 and 4-qutrit cases the optimal settings are mutually unbiased ($A=X$ and $A'=Z$) while in the 2 and 6-qutrit cases $A=X$ and $A'$ is a combination that includes $X$ \eqref{complexap}, so it cannot be unbiased with respect to $A$.

We have introduced the notion of {\sl multiplets of optimal settings} (MOS) which
 denotes any set of matrices that maximize the 2-qutrit and 6-qutrit inequalities,
  and all the 2 qudit inequalities. One is obtained from the other by applying a phase matrix and
 then the Fourier transform 
 and they have the property that both the commutator and the anticommutator of any pair of MOS are nilpotent matrices, i.e., matrices $M$ such that $M^k=0$ for some integer $k$. If one of the settings is set to $X$ then the other one has the following form
\[ MOS = e^{i\phi} \left( \begin{array}{ccccccc}
0 & 0 & 0 & ... & ... & ... & 1 \\
-1 & 0 & 0 & ... & ... & ... & 0 \\
0 & -1 & 0 & ... & ... & ... & 0 \\
...... & ... & ... & ... & ... & ... & ...\\
...... & ... & ... & ... & ... & ... & ...\\
...... & ... & ... & ... & ... & ... & ...\\
0 & ... & ... & ... & 0  & -1 & 0
\label{genmap}
\end{array} \right),\] 
where $\phi$ is a global phase that has to be tuned when changing between different forms of equivalent Bell inequalities. So, it is the same as $X$ but with opposite signs in all elements except for the first one, and a global phase.

\section{From two to three qutrits}
\label{2to3}

From the 2-qutrit inequality \eqref{cglmpim} it is possible to derive the 3-qutrit inequality \eqref{3qutrits}, under the assumption of symmetry for an additional third party. Starting from Eq.\eqref{cglmpim} it follows the sequence
\begin{eqnarray}
&& \hspace{2.5cm} [w (ab) - (a'b+ab') + w (a'b')]_A \leq \sqrt{3}, \nonumber\\ 
&& \hspace{1.8cm} [-i (w (ab) - (a'b+ab') + w (a'b'))]_H \leq \sqrt{3}, \nonumber\\
&&\hspace{1.0 cm}  [\frac{w^2-w}{\sqrt{3}} (w (ab) - (a'b+ab') + w (a'b'))]_H \leq \sqrt{3}, \nonumber\\
&&  [(1-w^2)(ab)+(w-w^2)(a'b+ab')+\nonumber\\
&&\hspace{4.7 cm} (1-w^2)(a'b'))]_H \leq 3, \nonumber\\
&&  [(ab)-w^2(ab+a'b+ab') + w(a'b+ab')+\nonumber\\
&&\hspace{5.0 cm} (w+2)(a'b')]_H \leq 3, \nonumber\\ 
&&  [(ab)-w^2(ab+a'b+ab') + w(a'b+ab'+a'b')+\nonumber\\
&&\hspace{6.0 cm} 2(a'b')]_H \leq 3.\nonumber\\
\end{eqnarray}
This form of the 2-qutrit CGLMP inequality suggests an 8-term symmetric inequality for three qutrits, where all terms with the same number of primes should have the same coefficients. By inserting $c$ and $c'$ according to this last requirement we have
\begin{eqnarray}
&[(abc)-w^2(abc'+a'bc+ab'c)+& \nonumber\\
&w(a'bc'+ab'c'+a'b'c)+2(a'b'c')]_H \leq 3.&
\end{eqnarray}
Thus, the symmetric 3-qutrit inequality \eqref{3qutrits} is obtained.

\section{Generalization to $d$ dimensions} \label{ddimensions}

In Ref. \cite{CollinsLinden} the bipartite CGLMP is extended to $d$ outcomes. Its expression in the probability language reads
\begin{eqnarray}
C_{22d} = && \sum_{k=0}^{[d/2]-1} \left( 1-\frac{2k}{d-1} \right)\\
&& \bigl(p(a=b+k)+p(b=a'+k+1)+\nonumber \\
&&p(a'=b'+k)+p(b'=a+k) \nonumber \\
&& -( p(a=b-k-1) + p(b=a'-k)+\nonumber \\
&&p(a'=b'-k-1) + p(b'=a-k-1))\bigr)\le 2. \nonumber
\end{eqnarray} 
Let us write these inequalities in term of operators. In order to do this let us start from a different form for \eqref{cglmpim} presented for example in Ref. \cite{Chen}
\begin{eqnarray}
C_{223} = &&  [ab+ab'+a'b-a'b']_H \nonumber \\
&& +\frac{1}{\sqrt{3}} [-ab+ab'+a'b-a'b']_A \le 2 \, .
\label{reim}
\end{eqnarray}     
In order to transform from probabilities to operators we have to establish a match between the number of variables and the number of equations. The variables here are the joint probabilities $p (a=b+k)$, with $k$ running from 0 to $d-1$, so there are $d$ unknowns. We need therefore $d$ equations. One equation is given by the normalization condition, i.e., the sum of probabilities is 1. For $d=2$, a second equation is enough, and that is the definition of expectation value of the product
\begin{equation}
ab = p(a=b)-p(a=b+1) \, .
\end{equation}
For $d=3$ there are 3 equations. Apart from the normalization of probabilities, two extra equations are needed, and those can be the hermitian and antihermitian parts of the expected value of the product, as in Eq. \eqref{reim}. It appears to be an accident that the CGLMP for $d=3$ can be expressed solely with the antihermitian part by inserting powers of $w$ as in Eq. \eqref{cglmpim}.

For $d=4$ we add the hermitian part of the expected values of the squares of products, and for $d=5$ we add their antihermitian part. The concrete expressions read as follows
\begin{eqnarray}\label{C42}
C_{224}&=&\frac{1}{3}\left( 2  [ab+ab'+a'b-a'b']_H \right.\\
      && + 2  [-ab+ab'+a'b-a'b']_A \nonumber \\
      && \left. +  [(ab)^2+(ab')^2+(a'b)^2-(a'b')^2]_H \right)\nonumber
\end{eqnarray}
and
\begin{eqnarray}\label{C52}
C_{225}&\!\!=\!\!&\frac{1}{2}\bigl( [ab+ab'+a'b-a'b']_H+\\
&&  [(ab)^2+(ab')^2+(a'b)^2-(a'b')^2 ]_H\bigr)+ \nonumber \\
&&\frac{2}{5} \bigl((3s_1+s_2)  [-ab+ab'+a'b-a'b']_A+ \nonumber \\
&&(-s_1+3s_2)  [-(ab)^2\!+\!(ab')^2\!+\!(a'b)^2\!-\!(a'b')^2 \bigr) ]_A\, ,\nonumber
\end{eqnarray} 
where the numbers $s_1$ and $s_2$ are the imaginary parts of $e^{2\pi i/5}$ and $e^{4\pi i/5}$, respectively. The classical bounds for these operators are $\langle C_{42}\rangle_{LR}=2$ and $\langle C_{52}\rangle_{LR}=2$.

It is possible to derive the general expression of the Bell operator for any number of levels $d$ as follows
\begin{equation}
C_{22d} = N \left( \sum_{k=1}^{[d/2]} r_{k,d}\mathrm{H}_{(ab)^k} + \sum_{k=1}^{[(d-1)/2]} i_{k,d} \mathrm{A}_{(ab)^k} \right) \le 2,
\end{equation}
where $r_{k,d}$ and $i_{k,d}$ are constants related to real and imaginary parts of $w$ (in general related to both of them), $N$ is a normalization constant such that the maximal classical value of $C_{22d}$ is 2, and also
\begin{eqnarray}
\textrm{H}_{(ab)^k}&\equiv&  [(ab)^k+(ab')^k+(a'b)^k-(a'b')^k ]_H, \nonumber \\
\textrm{A}_{(ab)^k}&\equiv&  [-(ab)^k+(ab')^k+(a'b)^k-(a'b')^k ]_A.\nonumber
\end{eqnarray}
All these inequalities are maximally violated by d-dimensional MOS as defined in Appendix \ref{mubs}. The numerical violation ratios increase with $d$, and can be found for example in Ref. \cite{Latorre}.

\end{document}